\documentclass[12pt,bibliography,preprint]{iopart}
\usepackage{graphicx}
\usepackage{epstopdf}
\usepackage{color}

\begin{document}
\title{Chiral symmetries associated with angular momentum}

\author{M. Bhattacharya}
\address{School of Physics and Astronomy, Rochester Institute of Technology, 84 Lomb Memorial Drive, Rochester, NY 14623, USA}
\ead{mxbsps@rit.edu}
\author{M. Kleinert}
\address{Department of Physics, Willamette University, Salem, OR 97301, USA}
\begin{abstract}
In beginning quantum mechanics courses, symmetries of a physical system are usually introduced as operators which commute with the Hamiltonian. In this article we will consider \textit{chiral} symmetries which \textit{anticommute} with the Hamiltonian. Typically, introductory courses at the (under)graduate level do not discuss these simple, useful and beautiful symmetries at all. The first time a student typically encounters them is when the Dirac equation is discussed in a course on relativistic quantum mechanics, or when particle-hole symmetry is studied in the context of superconductivity. In this article, we will show how chiral symmetries can be simply elucidated using the theory of angular momentum, which is taught in virtually all introductory quantum mechanics courses.
\end{abstract}
\vspace{2pc}

\section{Introduction}
``\textit{As far as I can see, all a priori statements in physics have their origin in symmetry}", said Hermann Weyl
\cite{WeylBook}. Symmetries play a paramount role in physics. They dictate the fundamental possibilities in a physical model,
including important features such as conservation laws, the nature of interactions, and selection rules for physical processes.
The search for new symmetries - and their destruction - thus forms a powerful guiding principle in the quest to understand
Nature \cite{GrossNobelLecture}. Last, but not least, the presence of symmetry adds to the aesthetic appeal of a physical
model and enhances its beauty.

In typical introductory courses at the undergraduate \cite{GriffithsBook} and graduate \cite{SchiffBook} levels,
symmetries are introduced as transformations that leave the Hamiltonian of a system unchanged. Mathematically, they
are represented by operators that commute with the Hamiltonian. Some common examples include symmetries under the
operations of translation, rotation, and inversion of spatial coordinates. System invariance under these symmetries implies
the existence of conserved quantities, such as linear momentum, angular momentum, and parity, respectively. The conservation
of such physical quantities inevitably simplifies, and often makes analytically solvable, the spectrum of the Hamiltonian and
the time evolution of the system \cite{SchiffBook}.

In this article, we consider symmetries that are represented by operators that \textit{anticommute} with the system
Hamiltonian. These `chiral' symmetries correspond to objects that are mirror images of each other. The best known
example of chiral symmetry is perhaps the case of the relativistic electron, where the charge conjugation operator anticommutes with the Dirac
Hamiltonian \cite{SchiffBook}. Another example from advanced textbooks is the Bogoliubov-de Gennes Hamiltonian in
superconductivity \cite{TinkhamBook}. Yet more specialized instances can be found in the research literature \cite{McIntosh1962}.
Recently, we analytically solved an effective Hamiltonian for the OH molecule by exploiting the presence
of chiral symmetry \cite{OH2013}.

While the existence of these symmetries cannot be associated with conserved quantities, they do simplify, and sometimes
render analytically solvable, the spectrum of the corresponding Hamiltonian. They also impose a rather pleasing
pattern on the energy spectrum, which students of physics may find appealing. In the present work we show that chiral
symmetries can be readily introduced and elaborated on at both the undergraduate and graduate levels, by simply
using the theory of angular momentum, which is taught in every introductory quantum mechanics course \cite{GriffithsBook}.

\section{Symmetries and chiral symmetries}
\label{sec:symm}
\subsection{Symmetries}
Typically, a symmetry of a physical system is mathematically represented as an operator $S$ that commutes with
the system Hamiltonian $H$, i.e.
\begin{equation}
\label{eq:commute}
[S,H] = SH-HS = 0.
\end{equation}
If the inverse $S^{-1}$ of the symmetry operator exists, then the second equality in Eq.~(\ref{eq:commute}) can be
rewritten as
\begin{equation}
\label{eq:invar}
S H S^{-1}=H,
\end{equation}
implying that the transformation under the symmetry operation $S$ leaves the Hamiltonian unchanged. The
existence of such a symmetry usually also implies the presence of degeneracy in the spectrum of the Hamiltonian $H$,
and often plays a crucial role in rendering the Hamiltonian analytically solvable. A familiar example is that
of the hydrogen atom, whose solvability relies partially on its invariance under rotations in three-dimensional space,
which in turn leads to the conservation of angular momentum \cite{SchiffBook}.
\subsection{Chiral symmetries}
An operator $C$ which anticommutes with the system Hamiltonian, i.e.
\begin{equation}
\label{eq:anticommute}
\{C,H\} = C H + H C =0,
\end{equation}
is said to correspond to a \textit{chiral} symmetry. The existence of such an operator implies that the spectrum
of the Hamiltonian consists of pairs of states with energies of equal magnitude, but opposite sign. This can be
seen readily by considering an eigenstate $\psi_{+}$ of $H$ with eigenvalue $+\lambda$,
\begin{equation}
\label{eq:plus}
H\psi_{+}=\lambda \psi_{+}.
\end{equation}
Multiplying both sides of Eq.~(\ref{eq:plus}) by $C$ and using Eq.~(\ref{eq:anticommute}) leads to
\begin{equation}
\label{eq:LHS}
C H\psi_{+}=C \lambda \psi_+=\lambda (C \psi_+) = -H \left( C \psi_{+}\right).
\end{equation}
This implies that
\begin{equation}
\psi_{-}= C\psi_{+},
\end{equation}
is an eigenstate of the Hamiltonian with eigenvalue $-\lambda$. Thus, the existence of the operator $C$ and its
anticommutation with $H$ imply the presence of paired energy levels $\pm \lambda$ in the spectrum as shown in Fig. \ref{fig:SP1} below. This phenomenon explains the term `chiral' symmetry since chiral comes from the Greek word for hand. Just like our left and right hand are mirror images of each other, the chiral pair in a spectrum consists of the group of positive and the group of negative energy states. These two groups are not identical, but are mirror images of each other across the zero-energy axis.

\subsection{Analytical solvability of finite dimensional Hamiltonians}
The eigenvalues of a Hamiltonian can be found by solving the equation $H = \lambda I$, where $H$ is the Hamiltonian, $\lambda$ the eigenvalue, and $I$ the unit matrix. This relation can be rewritten slightly by using the characteristic polynomial, $P(\lambda) = |H-\lambda I|$ and solving $P(\lambda)=0$ for $\lambda$. The analytical solubility of a finite-dimensional Hermitian matrix $H$ will generally depend on the dimension of the matrix. From the Abel-Ruffini theorem it is known that \textit{generally} only the characteristic polynomials of Hermitian matrices of dimension four or less can be solved using radicals \cite{StewartBook}, while those of dimension five can be solved using hypergeometric functions \cite{Green1978}. Polynomials of degrees higher than six are generally analytically unsolvable. Let us assume that the dimension
of the Hamiltonian matrix is $q$. The characteristic polynomial $P(\lambda)$ of the matrix $H$ is therefore also of degree $q$ in the variable $\lambda$, the eigenvalue. However, since the eigenvalues come in pairs for chiral symmetry, if $P(\lambda)=0,$ it follows
that $P(-\lambda)=0$ also. Thus, $P(\lambda)$ must be even in $\lambda$, unless $\lambda=0$ is a solution, in which case it can be factored out. $P(\lambda)$ can therefore be expressed as a polynomial of degree $q/2$ in the variable $\lambda^{2}.$ If $q/2 \leq 5,$ then the spectrum can be obtained analytically. This implies that chiral symmetry can enable the direct solution of even-dimensional Hamiltonian matrices of dimension up to $q=10.$ If the dimension of $H$ is odd, then there must be a zero energy state, which can be factored
out of $P(\lambda)$, leaving an even-dimensional polynomial whose roots can be found analytically. Specific examples will be provided below.
\section{Review of angular momentum algebra}
\label{sec:review}
For reference, we will summarize the elements of angular momentum theory required for our exposition of chiral symmetry \cite{GriffithsBook,SchiffBook}. We consider a total angular momentum operator $J$ with components $J_{x},J_{y},J_{z}$ such that
\begin{equation}
\label{eq:totalJ}
 J^{2}=J_{x}^{2}+J_{y}^{2}+J_{z}^{2}.
\end{equation}

\subsection{Commutation relations, basis and matrix representations}
\label{subsec:comm}
The commutation relations obeyed by the components of the angular momentum
are given by \cite{GriffithsBook}
\begin{equation}
[J_{x},J_{y}]=i\hbar J_{z}, \, [J_{y},J_{z}]=i\hbar J_{x},\, [J_{z},J_{x}]=i\hbar J_{y}.
\end{equation}
Further,
\begin{equation}
[J^{2},J_{x}]=[J^{2},J_{y}]=[J^{2},J_{z}]=0.
\end{equation}
A convenient basis is $|j,m \rangle$, where the eigenvalue $j$ is related to the magnitude of the total angular momentum,
\begin{equation}
J^{2}|j,m\rangle = j(j+1)\hbar^{2}|j,m \rangle,
\end{equation}
and the eigenvalue $m = -j, \ldots,0, \ldots, j$
\begin{equation}
\label{eq:jz}
J_{z}|j,m\rangle =m \hbar|j,m\rangle
\end{equation}
is the projection of the angular momentum vector onto a quantization axis.

Lastly, it is often convenient to introduce the raising and lowering operators $J_{\pm}$, defined by the following linear combinations of $J_x$ and $J_y$
\begin{equation}
J_{+}=J_x+i J_y,
J_{-}=J_x-i J_y.
\end{equation}

With these operators, we can write
\begin{equation}
J_{\pm}|j,m\rangle = \hbar \sqrt{j(j+1)-m(m \pm 1)}|j,m \pm 1\rangle,
\end{equation}
and we see that $J_+$ increases the $m$-value by one, while $J_-$ decreases it by one.

The matrix representation of the angular momentum operators can be constructed for any value of $j$ and $m$
from these relations. For example, for $j=1$ we can find the well known matrix representation
\begin{equation}
J_x = \frac{\hbar}{\sqrt{2}} \left( {\begin{array}{ccc}
0 & 1 & 0 \\
1 & 0 & 1 \\
0 & 1 & 0
\end{array} } \right),
J_y = \frac{\hbar}{\sqrt{2}i} \left( {\begin{array}{ccc}
0 & 1 & 0 \\
-1 & 0 & 1 \\
0 & -1 & 0
\end{array} } \right),
J_z = \hbar \left( {\begin{array}{ccc}
1 & 0 & 0 \\
0 & 0 & 0 \\
0 & 0 & -1
\end{array} } \right).
\end{equation}
The above commutation relations can be readily confirmed for these matrices.

\subsection{Rotations}
\begin{figure}
\includegraphics[width=0.6\textwidth]{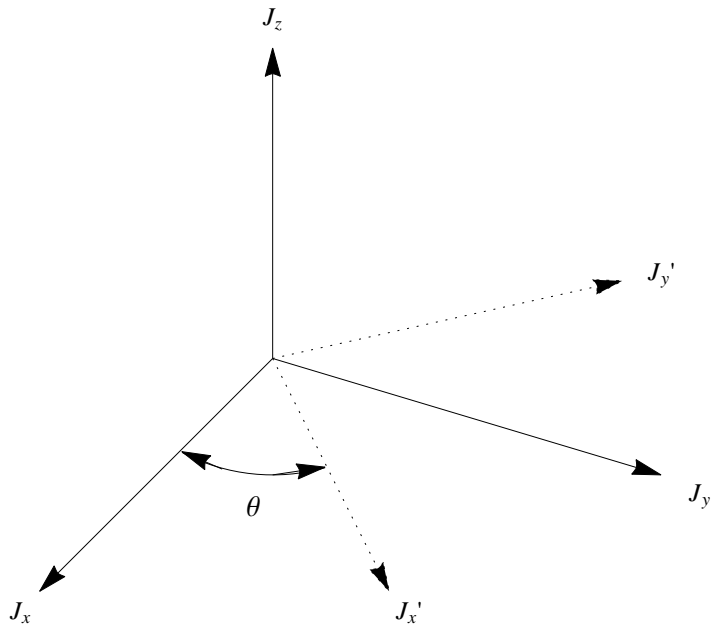}
\caption{The effect of the rotation operator $R_{z}(\theta)$ defined in Eq.~(\ref{eq:rotop}) on an operator is
a counter-clockwise rotation about the $z$-axis by the angle $\theta$. The angular momenta before and after the rotation
are shown in solid and dotted lines, respectively. For $\theta=\pi/2$, $J_{x}$ would transform to $J_y$, while $J_{y}$ would
rotate to $-J_{x}$.}
\label{fig:rot}
\end{figure}

It is well known that the angular momentum operators are the generators of rotation in quantum mechanics \cite{SchiffBook}.
The rotation operators are formed simply by exponentiating the operators of Section \ref{subsec:comm}.
For example, a counter-clockwise rotation by an angle $\theta$ about the $z$ axis can be represented by the operator
\begin{equation}
\label{eq:rotop}
R_{z}(\theta)=e^{-i\theta J_{z}/\hbar}.
\end{equation}
The effect of such the rotation $R_{z}(\theta)$ on any operator $M$ is given by $R_{z}(\theta)M R_{z}(\theta)^{-1}$, where $R(\theta)^{-1}=R(-\theta)$. An example is shown in Fig.~\ref{fig:rot}.

Here, we will focus on the special case where $M$ is an angular momentum operator, and the angle $\theta$ is such that the rotation yields the transformation $M \rightarrow -M.$ For example, we can easily show that
\begin{equation}
\label{eq:rot}
R_{z}(\pi)J_{x}R_{z}(-\pi)=-J_{x},
\end{equation}
which can be rewritten as
\begin{equation}
\label{eq:rot:anticom}
\{R_{z}(\pi),J_{x}\}=0.
\end{equation}

From Eq.~(\ref{eq:anticommute}) we know that the key to identifying chiral symmetry for a given Hamiltonian is to find a matrix that anticommutes with it. Mathematically, Eq.~(\ref{eq:rot:anticom}) thus implies chiral symmetry and sets the general strategy for the specific examples given in the following section. Physically, Eq.~(\ref{eq:rot:anticom}) is an example of the fact that a rotating object and its counter-rotating version are mirror-images of each other.

\section{Single spin in crossed magnetic fields}
\label{sec:singlespinsystems}
Our first physical example is a single spin in two crossed magnetic fields oriented along the $x$- and $y$-directions, respectively. Its Hamiltonian is given by
\begin{equation}
\label{eq:H1}
H_{1} = a J_{x}+b J_{y},
\end{equation}
where $a$ and $b$ are constants. It is straightforward to see that a rotation by $\pi$ about the $z$ axis changes the signs of both terms in $H_{1}$, see Eq.~(\ref{eq:rot}). Thus, $H$ anticommutes with the operator $R_{z}(\pi)$,
\begin{equation}
\{R_{z}(\pi),H_{1}\}=0.
\end{equation}

To illustrate this explicitly, let us consider the case  of a single electron in two crossed magnetic fields, i.e. $J=1/2$. In that case the angular momentum operators can be written in terms of the Pauli matrices, i.e. $J_{x}=\hbar \sigma_{x}/2=\hbar/2 \left( {\begin{array}{cc}
0 & 1  \\
1 & 0
\end{array} } \right),$
$J_{y}=\hbar \sigma_{y}/2=\hbar/2 \left( {\begin{array}{cc}
0 & -i  \\
i & 0
\end{array} } \right),$ and
$J_{z}=\hbar \sigma_{z}/2=\hbar/2 \left( {\begin{array}{cc}
1 & 0  \\
0 & -1
\end{array} } \right).$

Using these matrices, Eq.~(\ref{eq:H1}) can be written as
\begin{equation}
H_{1}(J=1/2)=\frac{\hbar}{2}\left(a\sigma_{x}+b\sigma_{y}\right) = \frac{\hbar}{2} \left({\begin{array}{cc}
0 & a-ib  \\
a+ib & 0
\end{array} } \right).
\end{equation}
This Hamiltonian is chiral symmetric since
\begin{equation}
\{R(\pi),H_1(J=1/2)\} = \left\{\left( {\begin{array}{cc}
-i & 0  \\
0 & i
\end{array} } \right),\frac{\hbar}{2} \left( {\begin{array}{cc}
0 & a-ib  \\
a+ib & 0
\end{array} } \right)\right\}=0.
\end{equation}
The chiral symmetry is also very obvious since the Hamiltonian can be diagonalized to yield the eigenvalues $\pm \hbar/2 \sqrt{a^2+b^2}$.
However, the anticommuting operator can also be chosen in this case to be simply $\sigma_{z},$ since the Pauli matrices all
anticommute with each other.

We note that the addition of the free rotational energy to Eq.~(\ref{eq:H1}),
\begin{equation}
\label{eq:H1P}
H_{1}'=a J_{x}+b J_{y}+c J^{2},
\end{equation}
where $a, b$ and $c$ are constants, makes the last term in Eq.~(\ref{eq:H1P}) a constant, $c J^{2}=c j(j+1)\hbar^{2}=C_{1}$, which simply shifts all the energies. This is because $[H_1',J^2]=0$. The spectrum is still reflection symmetric in this case, but about the nonzero energy $C_{1}$. In other words, the \textit{shifted} Hamiltonian $H_{1}'-C_{1}$ is chiral symmetric. We also note that the addition of odd powers such as $J_{x}^{3}$ to $H_{1}'$ will not destroy its chiral symmetry, while the addition of even powers such as $J_{x}^{2}$ will generally remove such a symmetry. This is because the rotation $R_{z}(\pi)$ anticommutes with the odd powers, but not with the even powers.

Let us now consider the more general case
\begin{equation}
\label{eq:HPP}
H_{1}''=a J_{x}+b J_{y}+c J_{z}.
\end{equation}
We can condense the notation and rewrite Eq.~(\ref{eq:HPP}) such that
\begin{equation}
H_{1}''= \textbf{a}\cdot \textbf{J},
\end{equation}
where $\textbf{a}=a\hat{x}+b\hat{y}+c\hat{z}$ is a vector and $\textbf{J}=J_{x}\hat{x}+J_{y}\hat{y}+J_{z}\hat{z}$ is a vector operator. Physically, the Hamiltonian $H_{1}''$ may then be thought of as the angular momentum operator along the direction of $\bf{a}.$ The rotation matrix Eq.~(\ref{eq:rotop}) can be generalized for a rotation by an angle $\theta$ about an axis denoted by a vector $\textbf{n}$ \cite{SchiffBook}
\begin{equation}
R_{\textbf{n}}(\theta)=e^{-i\theta \textbf{n}\cdot \textbf{J}/\hbar}.
\end{equation}

We can consider an identity that reveals the effect of this rotation on $H_{1}''$:
\begin{equation}
\label{eq:genrot}
R_{\textbf{n}}(\theta)\textbf{a}\cdot \textbf{J}R_{\textbf{n}}(\theta)^{-1}
=\cos\theta\textbf{a}\cdot \textbf{J}+\sin\theta\left(\textbf{n}\times \textbf{a}\right)\cdot \textbf{J}
+\left(1-\cos\theta\right)\left(\textbf{n}\cdot\textbf{J}\right)\left(\textbf{n}\cdot\textbf{a}\right).
\end{equation}
The reader is encouraged to follow the hint supplied in Merzbacher \cite{MerzbacherBook} to prove this
identity. The proof itself can be found in \cite{PuriBook}, for example. Examining Eq.~(\ref{eq:genrot}) by picking specific directions for
the vector \textbf{a} such as $\hat{x}$, etc. is also a useful exercise. Particularly, we realize that by choosing $\textbf{n}\cdot\textbf{a}=0$ and $\theta=\pi$, Eq.~(\ref{eq:genrot})
reads
\begin{equation}
R_{\textbf{n}}(\pi)\textbf{a}\cdot \textbf{J}R_{\textbf{n}}(\pi)^{-1}
=-\textbf{a}\cdot \textbf{J},
\end{equation}
which implies the presence of chiral symmetry, as then
\begin{equation}
\{R_{\textbf{n}}(\pi), H_{1}''\}=0.
\end{equation}
To illustrate, let us consider two specific examples. First, let us calculate $H_{1}''$ for $J=1,$ a system the student is probably
very familiar with, which yields a $2J+1=3$-dimensional Hermitian matrix
\begin{equation}
\label{eq:H1PPMat}
H_{1}''=\left(
  \begin{array}{ccc}
   c & \frac{1}{\sqrt{2}}\left( a - i b\right) & 0  \\
    \frac{1}{\sqrt{2}}\left( a + i b\right) &  0 &\frac{1}{\sqrt{2}}\left( a - i b\right) \\
    0 & \frac{1}{\sqrt{2}}\left( a + i b\right) & -c \\
   \end{array}
\right).
\end{equation}

The characteristic polynomial of this matrix is given by
\begin{equation}
\label{eq:CPH1PP}
P(\lambda)=\lambda\left(-\lambda^{2}+\hbar^2 Q\right),
\end{equation}
where $Q=(a^{2}+b^{2}+c^{2}).$ Note that the even powers of $\lambda$ are absent from Eq.~(\ref{eq:CPH1PP}); however, one power of $\lambda$ can be factored, as shown.
The eigenvalues of the matrix in Eq.~(\ref{eq:H1PPMat}) can be readily found by setting the characteristic polynomial equal to
zero. This leads to eigenvalues $0, \pm \hbar \sqrt{Q}.$ Clearly, the spectrum has reflection symmetry about zero energy.

\begin{figure}
\includegraphics[width=0.8\textwidth]{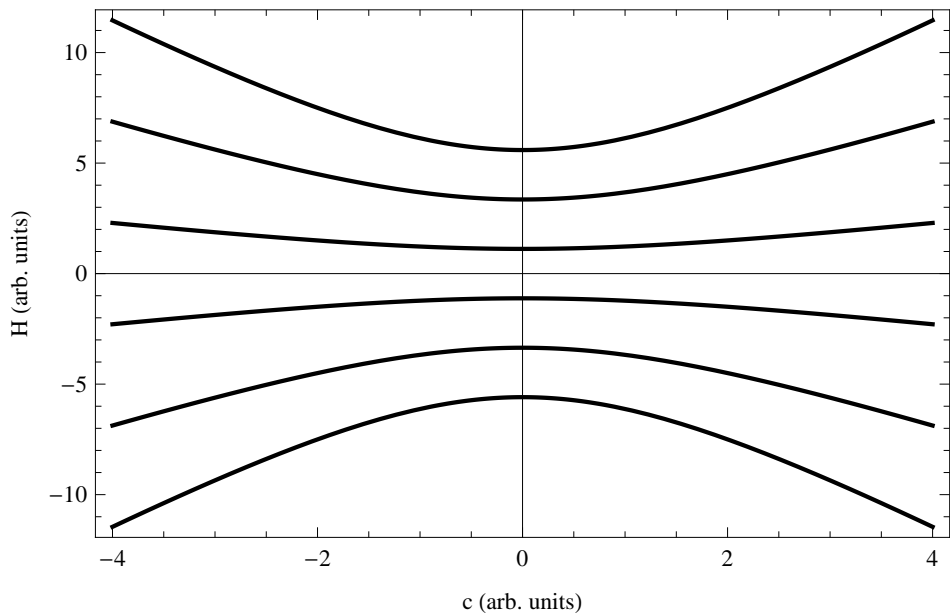}
\caption{Spectrum of the Hamiltonian $H_{1}''$ of Eq.~(\ref{eq:H1PPMat2}) plotted as a function of the parameter
$c$ with $a=1,b=2$.}
\label{fig:SP1}
\end{figure}

A more interesting and less trivial system is the case of $J=5/2,$ which yields a $2J+1=6$-dimensional Hermitian matrix
\begin{equation}
\label{eq:H1PPMat2}
H_{1}''=\frac{1}{2} \left(
  \begin{array}{cccccc}
    5c & \sqrt{5}\left( a - i b\right) & 0 & 0 & 0 & 0 \\
    \sqrt{5}\left( a + i b\right) & 3c &  \sqrt{8}\left( a - i b\right) & 0  & 0 & 0 \\
    0 & \sqrt{8}\left( a + i b\right) & c & 3 \left(a-i b\right) & 0 & 0 \\
    0 & 0 & 3 \left(a+i b\right)  & -c & \sqrt{8}\left( a - i b\right) & 0 \\
    0 & 0 & 0 & \sqrt{8}\left( a + i b\right) & -3c & \sqrt{5}\left( a - i b\right) \\
    0 & 0 & 0 & 0 & \sqrt{5}\left( a + i b\right) & -5c  \\
  \end{array}
\right).
\end{equation}
It is not \textit{a priori} evident that this Hamiltonian can be solved analytically. But because of the chiral symmetry, its characteristic polynomial is a cubic in $\lambda^2$
\begin{equation}
\label{eq:CPH1PP2}
P(\lambda)=\left(\lambda^{2}\right)^{3}+c_{4}\left(\lambda^{2}\right)^{2}+c_{2}\lambda^{2}+c_{0},
\end{equation}
which can be solved easily. The coefficients are given by
\begin{equation}
c_{0}=-\frac{225 Q^{3}}{64}, \, c_{2}=\frac{259 Q^{2}}{16}, \, c_{4}=-\frac{35 Q}{4}.
\end{equation}
Note that the odd powers of $\lambda$ are absent from Eq.~(\ref{eq:CPH1PP2}).
The eigenvalues of the matrix in Eq.~(\ref{eq:H1PPMat2}) can be readily found to be
$\pm 5\sqrt{Q}/2, \pm 3\sqrt{Q}/2, \pm \sqrt{Q}/2.$ They are plotted in Fig.~\ref{fig:SP1} for $a=1, b=2$
and various values of $c$. Clearly, the spectrum has reflection symmetry about zero energy.

\section{Triaxial rotor}
Another interesting physical example is the atomic nucleus which can be modeled as a spin that is allowed to rotate about all three axes with moments of inertia $(I_{x},I_{y}, I_{z}) > 0$, respectively \cite{Wood2004}. It can be described by the Hamiltonian
\begin{equation}
\label{eq:H2}
H_{2}= \frac{J_{x}^{2}}{2I_{x}}+\frac{J_{y}^{2}}{2I_{y}}+\frac{J_{z}^{2}}{2I_{z}}.
\end{equation}
It may seem that there cannot be chiral symmetry in this case as $H_{2}$ is even in all components of the angular momentum. However, chiral symmetry can exist for some special values of the moments of inertia. It is important to note that using
Eq.~(\ref{eq:totalJ}), we can rewrite Eq.~(\ref{eq:H2}) as
\begin{equation}
\label{eq:H22}
H_{2}= \left(\frac{1}{2I_{x}}-\frac{1}{2I_{z}}\right)J_{x}^{2}+\left(\frac{1}{2I_{y}}-\frac{1}{2I_{z}}\right)J_{y}^{2}+\frac{J^{2}}{2I_{z}}.
\end{equation}
Note that $[H_{2},J^{2}]=0,$ and similar to the case of Eq.~(\ref{eq:H1P}), the spectrum will simply be offset
by the constant term
\begin{equation}
J^{2}/2I_{z}=j(j+1)\hbar^{2}/2I_{z}=C_{2}.
\end{equation}
Now consider the special situation where
\begin{equation}
\frac{1}{I_{x}}+\frac{1}{I_{y}}=\frac{2}{I_{z}},
\end{equation}
which implies that
\begin{equation}
\left(\frac{1}{2I_{x}}-\frac{1}{2I_{z}}\right)=-\left(\frac{1}{2I_{y}}-\frac{1}{2I_{z}}\right)=D.
\end{equation}
Using this relation in Eq.~(\ref{eq:H22}), we may write the shifted Hamiltonian
\begin{equation}
\label{eq:H2P}
H_{2}'=H_{2}-C_{2}=D(J_{x}^{2}-J_{y}^{2}).
\end{equation}
A little thought reveals that this Hamiltonian has chiral symmetry, since a rotation by $\pi/2$ about the
$z$ axis changes $J_{x}\rightarrow  J_{y}$ and $J_{y}\rightarrow - J_{x},$ see Fig.~\ref{fig:rot}. Since this transformation
exchanges the $x$ and $y$ labels on the angular momentum components in Eq.~(\ref{eq:H2P}), it reverses the
sign of $H_{2}'$. Indeed, $\{R_{z}(\pi/2),H_{2}'\}=0.$

This example shows that in order for chiral symmetry to be present in a system, it is not necessary that each individual term in the Hamiltonian be chiral symmetric.

\section{Two coupled spins}
Finally, we will consider a system of two coupled spins of angular momenta $J_{1}$ and $J_{2}$, respectively. Finding the rotations that anticommute with the Hamiltonian becomes only a little more complicated. We will first present a toy model to illustrate the method. Then we will discuss a real system, namely the ground state OH molecule in combined electric and magnetic fields, that we have recently solved analytically \cite{OH2013}.

\subsection{Toy model}
Let us consider the toy Hamiltonian
\begin{equation}
\label{eq:H3}
H_{3} = A J_{1,y}J_{2,y}+B J_{1,z}J_{2,z},
\end{equation}
where $A$ and $B$ are constants. This Hamiltonian describes two spins that are coupled along the $y$ and $z$ directions, but not along $x$. Since we are now considering two spins, the anticommuting rotation operator must be a product of rotation operators of the individual spins. Also, these operators cannot correspond to rotations
about the same axis, as then each term in Eq.~(\ref{eq:H3}) would pick up two negative signs, one from each spin, and would overall remain unchanged. Thus, we must consider rotating each spin about a different axis.

After some tinkering we find that the correct rotation operator that anticommutes with $H_{3}$ is given by
\begin{equation}
R_{3}=R_{1,y}(\pi)R_{2,z}(\pi).
\end{equation}
This operator rotates the first spin by $\pi$ about the $y$ axis and the second spin about the $z$ axis by the same angle. Looking at the first term in Eq.~(\ref{eq:H3}), we see that $R_3$ leaves the first spin unchanged but rotates the second spin by $\pi$ about the $z$ axis. Thus, it anticommutes with $J_{2,y}$ and thus with the full first term in Eq.~(\ref{eq:H3}). Similar reasoning shows that $R_{3}$ anticommutes with the second term as well. Thus,
\begin{equation}
\{R_{3},H_{3}\}=0.
\end{equation}

\subsection{Ground state OH molecule in combined electric and magnetic fields}
\begin{figure}
\includegraphics[width=0.8\textwidth]{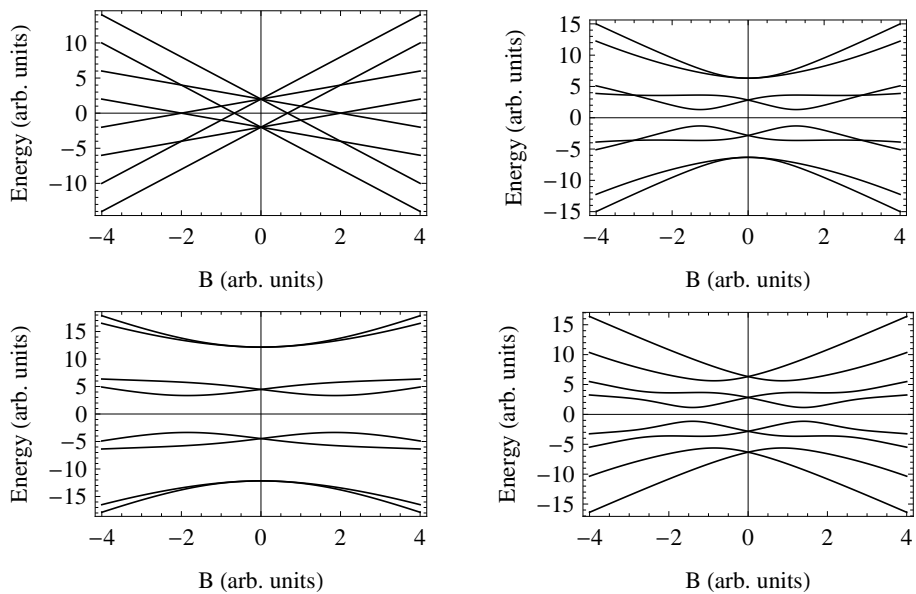}
\caption{Spectra of the Hamiltonian $H_{4}$ describing the OH molecule, plotted as a function of the parameter $B$, for various values
of $\Delta,E$ and $\theta$. Note that although the pattern of crossings and avoided crossings changes dramatically, chiral symmetry is
always maintained. Also note that the exact values of $\Delta$, $E,$ and $\theta$ are not relevant to this discussion. The reader is referred to Ref.~\cite{OH2013} for a more quantitative discussion of the spectrum.}
\label{fig:OHSpec}
\end{figure}

Consider now a Hamiltonian that describes the ground state OH molecule in combined electric and magnetic fields \cite{OH2013}
\begin{equation}
\label{eq:OH}
H_{4}=\Delta J_{1,z}+ B J_{2,z}+ E J_{1,x}(J_{2,z}\cos\theta-J_{2,x}\sin\theta),
\end{equation}
where $\Delta,B,E$ and $\theta$ are constants related to the internal structure of the molecule, the magnetic field, the electric field and the angle between the two fields, respectively. For the OH molecule $J_{1}=1/2$ and $J_{2}=3/2.$ An appropriate basis in that case can be
constructed from the product states $|j_{1},m_{1}\rangle|j_{2},m_{2}\rangle$ where $m_{1}$ takes values $-1/2, 1/2$ and $m_{2}$ takes values $-3/2, -1/2,1/2,3/2$. The basis thus has dimension $(2j_{1}+1)(2j_{2}+1)=2\times 4=8.$ The matrix elements of the operator products in Eq.~(\ref{eq:OH}) can be evaluated in this basis. For example
\begin{eqnarray}
\langle j_{1}',m_{1}'|\langle j_{2}',m_{2}'|J_{1,x}J_{2,z}|j_{1},m_{1}\rangle |j_{2},m_{2}\rangle \nonumber\\
=\langle j_{1}',m_{1}'|J_{1,x}|j_{1},m_{1}\rangle \langle j_{2}',m_{2}'|J_{2,z}|j_{2},m_{2}\rangle,\\
\nonumber
\end{eqnarray}
where the matrix elements of the first and second spins can be individually evaluated using the relations in Section~\ref{subsec:comm}.
Typical spectra that clearly show chiral symmetry are shown in Fig.~\ref{fig:OHSpec}.

The anticommuting rotation operator may now be reasoned out in the following way: The first two terms in Eq.~(\ref{eq:OH}) contain spins that are coupled along the $z$ direction. Thus, in order to change the sign of the first or the second term, we need to rotate the first or second spin by $\pi$ about either the $x$ or the $y$ axis. However, this means that $J_{2,z}$ in the third term will also change sign. Thus, we should rotate the second spin in such a way that $J_{2,x}$ changes sign as well. This can be accomplished by picking the rotation operator $R_{2,y}(\pi).$ Now we need to ensure that the $J_{1,x}$ does \textit{not} change sign as $J_{1,z}$ reverses. This can be done by choosing the rotation operator $R_{1,x}(\pi)$. Thus, our final rotation operator is given
by
\begin{equation}
R_{4}=R_{1,x}(\pi)R_{2,y}(\pi),
\end{equation}
such that
\begin{equation}
\{R_{4},H_{4}\}=0.
\end{equation}
The matrix representation of the Hamiltonian $H_{4}$ is of dimension $(2J_{1}+1)(2J_{2}+1)=8.$ Chiral symmetry reduces this effectively to a
quartic, making it exactly solvable. The analytic expressions for the energy eigenvalues can be found in
Ref.~\cite{OH2013}.

Before we conclude, we would like to note that since the angular momentum operators are Hermitian, rotation
operators are unitary. The inverse $R^{-1}$ of a rotation operator $R$ is given by its Hermitian conjugate, and is
well-defined. In this case, the anticommutation $R H+H R=0$ can be rephrased as $RHR^{-1}=-H,$ which contrasts
nicely with Eq.~(\ref{eq:invar}). However, rather than start with this similarity transformation, we have chosen
to make the anticommutation relation as our fundamental requirement for the presence of chiral symmetry. This was
done to keep the discussion general. Specifically, our approach includes operators which anticommute with the
Hamiltonian but whose inverse may not exist.

As we have shown, chiral symmetry of a Hamiltonian implies reflection symmetry in the spectrum. One may think of this reflection
symmetry as a rotation of the spectrum itself by $\pi$. But what if a spectrum returns to itself when rotated by
$\pi/n,$ where $n$ is an integer? This more general kind of rotational symmetry has been addressed mathematically
by the Perron-Frobenius theorem \cite{BermanBook}. However, this theorem applies only to matrices whose elements are all non-negative.
It may be verified readily that matrices representing angular momentum operators do not fall in this category.

\section{Discussion}
\label{sec:discussion}
In this article we have shown that chiral symmetries can be accessibly incorporated in undergraduate
and graduate quantum mechanics courses using the basic theory of angular momentum. These chiral symmetries can be
represented by operators which anticommute with the Hamiltonian. While these symmetries do not correspond to
the existence of conserved quantities, we have shown that they simplify and often render analytically solvable
the spectrum of the Hamiltonian.

As in the case of the more familiar symmetries, there seems to be no systematic method
for finding chiral symmetries. The accomplishment of this task therefore requires intuition developed by solving
specific examples. We have attempted to supply such examples in this article. We hope that these examples will
prepare and familiarize students with chiral symmetries before they encounter them in more advanced courses such as
relativistic quantum mechanics and superconductivity.\\

\textbf{References}

\end{document}